`\pdfoutput=1`
\documentclass[prl,preprintnumbers,tightenlines,twocolumn,superscriptaddress]{revtex4-1}
\usepackage{braket}
\usepackage{dsfont}
\usepackage{amsmath}
\usepackage{amssymb}
\usepackage{amsfonts}
\usepackage{mathtools}
\usepackage{graphicx}
\usepackage{dcolumn}
\usepackage{bm}
\usepackage{multirow}
\usepackage{leftidx}
\usepackage{color}
\usepackage{mathtools}
\usepackage{MnSymbol}
\usepackage[mathscr]{eucal}
\usepackage[german,english]{babel}
\usepackage[capitalize]{cleveref}
\usepackage[utf8]{inputenc}

\makeatletter
\DeclareRobustCommand{\cev}[1]{%
  \mathpalette\do@cev{#1}%
}
\newcommand{\do@cev}[2]{%
  \fix@cev{#1}{+}%
  \reflectbox{$\m@th#1\vec{\reflectbox{$\fix@cev{#1}{-}\m@th#1#2\fix@cev{#1}{+}$}}$}%
  \fix@cev{#1}{-}%
}
\newcommand{\fix@cev}[2]{%
  \ifx#1\displaystyle
    \mkern#23mu
  \else
    \ifx#1\textstyle
      \mkern#23mu
    \else
      \ifx#1\scriptstyle
        \mkern#22mu
      \else
        \mkern#22mu
      \fi
    \fi
  \fi
}

\makeatother

\newcommand{\dd}{\, \mathrm{d}}

\newcommand{\be}{\begin{equation}}
\newcommand{\ee}{\end{equation}}

\newcommand{\E}[1][\empty]{
  \ifthenelse{\equal{#1}{\empty}}
    {\mathbb{E}}
    {\mathbb{E}\left( #1 \right)}
}
\renewcommand{\exp}[1][\empty]{
  \ifthenelse{\equal{#1}{\empty}}
    {\mathrm{exp}}
    {\mathrm{e}^{#1}}
}

\newcommand{\psit}[1][\empty]{%
  \ifthenelse{\equal{#1}{\empty}}
    {\psi_t}
    {\psi_t^{(#1)}}
}

\newcommand{\npsit}[1][\empty]{%
  \ifthenelse{\equal{#1}{\empty}}
    {\tilde\psi_t}
    {\tilde\psi_t^{(#1)}}
}

\usepackage{ulem}  
\newcommand{\oalex}[1]{{\color{blue}{}}}

\definecolor{olive}{RGB}{107,142,35}

\definecolor{orange}{RGB}{255,139,61}

\newcommand{\om}[1]{{}}

\graphicspath{{{./}}}
\begin{document}
\title{Dressed ion-pair states of an ultralong-range Rydberg molecule}
\author{P. Giannakeas}
\email{pgiannak@pks.mpg.de}
\affiliation{Max-Planck-Institut f\"ur Physik komplexer Systeme, N\"othnitzer Str.\ 38, D-01187 Dresden, Germany }
	
\author{Matthew T. Eiles}
\email{meiles@pks.mpg.de}
\affiliation{Max-Planck-Institut f\"ur Physik komplexer Systeme, N\"othnitzer Str.\ 38, D-01187 Dresden, Germany }

\author{F. Robicheaux}
\email{robichf@purdue.edu}
\affiliation{Department of Physics and Astronomy, Purdue University, 47907 West Lafayette, IN, USA}
\affiliation{ Purdue Quantum Science and Engineering Institute, Purdue University, West Lafayette, Indiana 47907, USA}
\author{Jan M. Rost}
\email{rost@pks.mpg.de}
\affiliation{Max-Planck-Institut f\"ur Physik komplexer Systeme, N\"othnitzer Str.\ 38, D-01187 Dresden, Germany }

\begin{abstract}
We predict the existence of a universal class of ultralong-range Rydberg molecular states whose vibrational spectra form trimmed Rydberg series.
A dressed ion-pair model captures the physical origin of these exotic molecules, accurately predicts their properties, and reveals features of ultralong-range Rydberg molecules and heavy Rydberg states with a surprisingly small Rydberg constant.
The latter is determined by the small effective charge of the dressed anion, which outweighs the contribution of the molecule's large reduced mass. This renders these molecules the only known few-body systems to have a Rydberg constant smaller than $R_\infty/2$.
\end{abstract}
\maketitle
The richness of Rydberg physics is highlighted by two exotic molecular systems which have attracted recent interest: ultralong-range Rydberg molecules (ULRM) and heavy Rydberg states (HRS).
The ULRM is a fragile dimer with a bond length $\sim$100 nanometers.
This gargantuan molecule consists of a neutral perturber atom ($B$) bound to a highly excited Rydberg atom ($A^*$)  \cite{Greene2000,HosseinReview,EilesTutorial,HamburgReview}.
The experimental observation of ULRMs \cite{bendkowsky,TallantCS,DeSalvo2015} has led to their use in many diverse applications, e.g. as probes of charge-neutral interactions \cite{SpinFey,PhasesFeyMeinert,Sass,MacLennan,IonColdMeinert,NewPfau} or as impurities embedded in a many-body bath \cite{WhalenFermions,Schlag,UltracoldChem,MolSpec,IonRydBlockade,WhalenSpatial, Demler,WhalenPoly,Whalen2,Ashida1,Sous1}.
ULRMs exist because the Rydberg electron accumulates an appreciable phase shift as it scatters off of the perturber, which in turn produces an energy shift proportional to the $S$-wave scattering length \cite{Fermi,Omont}.
This binds the ``trilobite'' molecule, $A^*B$, together \cite{PfauRaithel,BoothTrilobite,PfauSci}.
 If the electron-perturber ($e-B$) complex possesses a $P$-wave shape resonance, a second, more deeply bound, "butterfly" ULRM forms  \cite{HamiltonGreeneSadeghpour,KhuskivadzeJPB,butterfly}.

HRS (also called ion pairs or heavy Bohr atoms) are the direct molecular analogues of a Rydberg atom \cite{Ubachs1,Ubachs2,Ubachs3,HRS,MerktHRS}.
An atomic anion replaces the Rydberg electron, creating a vibrational Rydberg state $A^+ B^-$.
The properties of these molecules obey typical Rydberg formulae, but with the electron's mass replaced by the dimer's heavy reduced mass.
Typically the excitation of pairs of ground state atoms to HRS is difficult due to weak Franck-Condon factors and electronic transition-dipole moments \cite{HRS,HRS2}.
Recent proposals exploit ULRMs, with similar bond lengths as HRS, to avoid these challenges \cite{DeiglmayrHRS,HummelHRS}.
 In the vicinity of the perturber, the electronic wavefunction of the butterfly molecule and the metastable excited $P$ anion have the same symmetry \cite{PhasesFeyMeinert}.
 The electron can thus be transferred from the Rydberg state into the bound $S$ anion state via a dipole-allowed transition, which also supplies the required energy to match the electron affinity and allow the reaction $A^*B \to A^+ B^-$ to occur.

\begin{figure*}[t]
	\includegraphics[width=1.99\columnwidth]{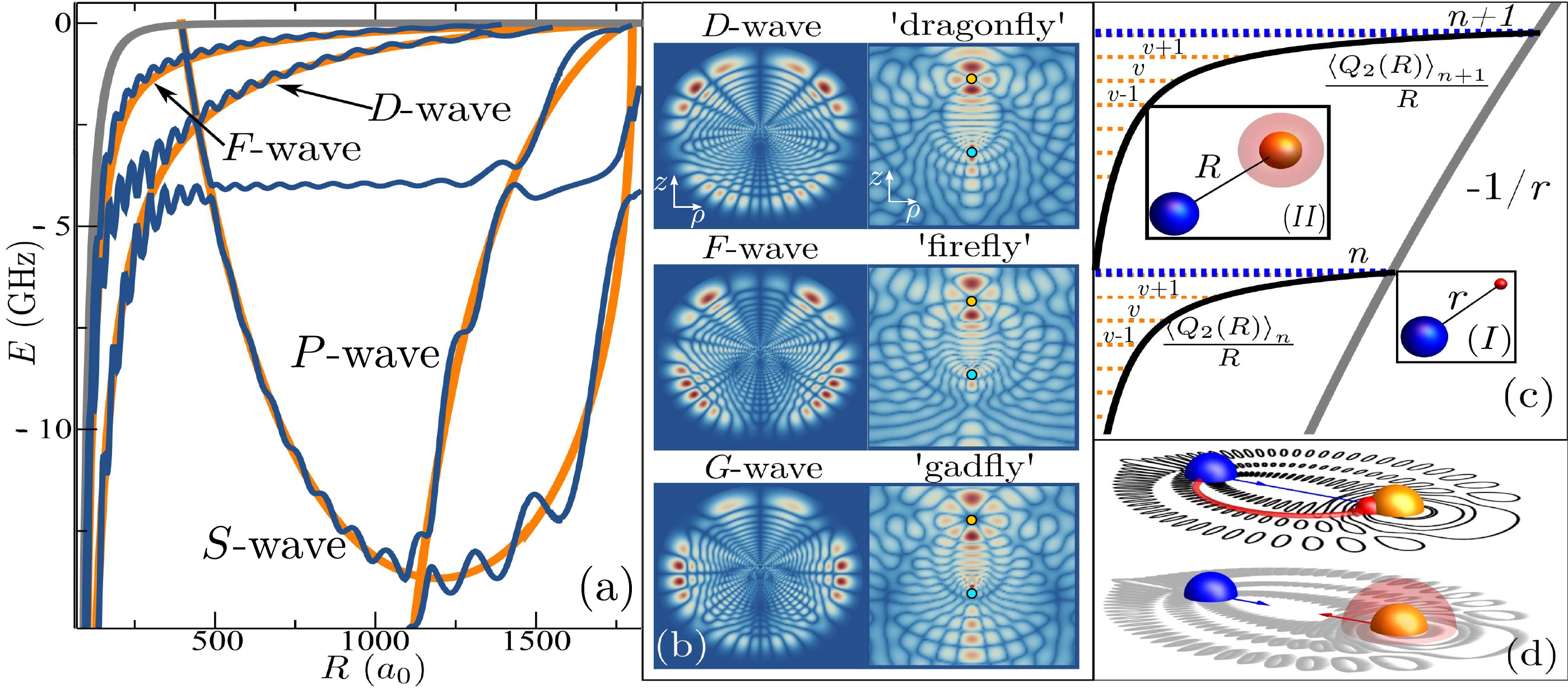}
	\caption{(color online) (a) $^\Sigma$Rb$_2$ Rydberg molecule PECs (blue) relative to the $n=30$ manifold. The smooth orange curves overlaid are the results of the model, \cref{eq:pec}, and the ion-atom polarization potential is shown in gray.
(b) Electronic wavefunctions $\Psi_\Sigma(z,\rho)$ in cylindrical coordinates are depicted for the family of $L\ge 2$ states.
The left column shows $\rho\sqrt{| \Psi_\Sigma(z,\rho)|}$ for $z \in[-2000,2000]$ and $\rho \in[-2000,2000]$.
The right column shows $\sqrt{|\Psi_\Sigma(z,\rho)|}$ for $z \in[-250,300]$ and $\rho \in[-250,250]$.
The blue (orange) dot indicates the position of the Rydberg core (perturber).
(c) A schematic of the vibrational Rydberg series. The electron-ion potential (gray) supports the bound states of the Rydberg atom (I). Each electronic Rydberg manifold labeled $n$, $n+1\dots$ (blue dashed lines) is the threshold for the $M=2$ $D$-wave PEC (black).
This PEC supports a nuclear Rydberg series (orange dotted lines) of dressed ion-pair states (II) labeled $v-1$, $v\dots$.
(d) Two pictures of a ULRM: On top, the typical picture, where the Rydberg electron (red) scatters from the perturber (orange) and forms a trilobite-like wave function (black contour).
On the bottom, the dressed ion-pair picture, wherein the Rydberg wavefunction is neglected and the relevant quantity is the electron's charge distribution (red) around the perturber, forming a dressed anion.
}
\label{fig:fig1}
\end{figure*}

 However, the exotic systems of ULRMs and HRS exist typically in well separated energy intervals.
 In this Letter, we predict a class of highly excited molecular states which realize properties of HRS on the electronic energy scale of ULRMs, thus combining both concepts.
The inclusion of higher partial waves ($L\ge 2$) in the $e-B$ interaction yields a hierarchy of ``truncated Coulomb'' potential energy curves (PECs) governing the \textit{vibrational} motion associated with every degenerate \textit{electronic} Rydberg manifold, labeled by $n$.
Each level in the infinite electronic Rydberg series becomes the dissociation threshold for a set of trimmed heavy Rydberg series, or tHRS (see Fig.~\ref{fig:fig1}(c)).
The tHRS possess only a finite number of vibrational states since the ``truncated Coulomb'' PECs vanish once the perturber resides outside the Rydberg electron's orbit.
The tHRS preserve the basic attributes of ULRMs; in particular, the perturbed electronic wavefunction fills the entire Rydberg volume, while in HRS the electron binds to the perturber.
To address the physical origin of the tHRS, we map the ULRM system, $A^*B$, (upper panel of Fig.~\ref{fig:fig1}(d)) onto an effective ion-pair, $A^+B^{Q}$,  (lower panel of Fig.~\ref{fig:fig1}(d)) where the perturber is {\it dressed by a fractional charge $Q$}.

The dressed ion-pair model reveals that the perturber-induced localization of the electron yields to an effective negative charge which is virtually independent of the internuclear distance $R$, leading thus to Coulomb-like molecular PECs.
This property arises due to the linear energy dependence of the electron-perturber phase shifts for $L\ge2$ which is universally satisfied for {\it any} type of polarizable perturber at low energies.
Additionally, the magnitude of the fractional charge counterbalances the heavy mass of the tHRS and leads to a Rydberg constant which is significantly smaller than $R_\infty/2$, the Rydberg constant of positronium, the lightest ion pair. 

We compute the ULRM PECs including $L\ge 2$ phase shifts by employing the generalized local frame transformation (GLFT) approach \cite{pra_panos}.
This framework requires as input only a set of atomic quantum defects $\mu_l$ and scattering phase shifts $\delta_L$ with $l$ ($L)$ indicating the electron's angular momentum relative to $A^+$ ($B$).
The generic scope of the GLFT approach permits us to include the effect of higher partial wave e-B scattering, avoiding the limitations of the Fermi/Omont pseudopotential or Green's function methods \cite{Fermi,Omont,EilesTutorial,EilesSpin,Fey,Crowell,HamiltonThesis,KhuskivadzePRA}.
The triplet $e-B$ scattering phase shifts are obtained from a non-relativistic two-electron $R$-matrix code \cite{TaranaCurikLi,EilesHetero}.
We consider as a paradigm spinless Rb atoms and use atomic units unless otherwise specified.
Thus, $M$, the projection of $L$ onto the internuclear axis $R$, is a good quantum number and defines the molecular symmetry.
Focusing on the universal aspects of the tHRS, realizable in spin-stretched experiments, we neglect relativistic effects and singlet states.

\cref{fig:fig1}(a) shows the $\Sigma$ $(M=0)$ $\rm{Rb}_2$ ULRM PECs relative to the $n=30$ Rydberg manifold including $L\le 3$ partial waves.
The blue curves depict the trilobite PEC ($S$-wave scattering), the butterfly PEC ($P$-wave), as well as a new series of $L\ge 2$ PECs which complete the ULRM ``family'': the ``dragonfly'' PEC ($D$-wave), ``firefly'' PEC ($F$-wave), etc.
As $L$ increases the PECs condense into the ion-atom potential $-\frac{\alpha}{2R^4}$ (gray curve), where $\alpha$ is the atomic polarizability.
Exemplary electronic wavefunctions with the perturber placed at $R=200~a_0$ are displayed in \cref{fig:fig1}(b).
Near the perturber they manifest the dominant spherical symmetry of the $e-B$ interaction since the Coulomb field is negligible.
Thus, $L$ is approximately a good quantum number labeling these states.
These molecules exhibit dipole moments in the hundreds of debye, similar to the trilobite/butterfly molecules [see Supplemental material (SM)].

Unlike the trilobite and butterfly PECs, the $L\ge 2$ PECs are, to a good approximation, Coulombic. This is particularly evident for molecular states having higher $M$ values, since the oscillatory fringes in the PECs vanish for increasing $M$.
For example, \cref{fig:fig1}(c) schematically depicts in black the $\Delta$ ($M=2$) dragonfly PECs which dissociate at each electronic Rydberg $n-$manifold (blue dotted lines).

The dressed ion-pair model captures intuitively the emergence of the Coulombic character in the $L\ge 2$ PECs, simultaneously illustrating why the $S$ and $P$-wave PECs are so different.
The standard depiction of an ULRM, for example a trilobite, is depicted in the upper panel of \cref{fig:fig1}(d).
The perturber mixes the degenerate Rydberg states to form the trilobite wavefunction, plotted as a contour.
The nodal pattern of this electronic wavefunction is linked to the oscillatory fringes of the $S-$wave PEC [see \cref{fig:fig1}(a)].
Starting out tabula rasa, in the bottom panel of \cref{fig:fig1}(d) the ULRM is viewed as an effective ion-pair where the perturber is dressed by a charge distribution (red sphere in \cref{fig:fig1}(d)) forming an anion of fractional charge.
This perspective ignores the total trilobite wavefunction except for the charge distribution localized by the $e-B$ interaction.
  
 The effective charge is obtained by calculating the difference in electronic probability accumulated in the vicinity of the perturber with and without its presence.
 The resulting integral is evaluated in terms of the phase shift $\delta_L(k)$ at a given electronic scattering kinetic energy $\frac{(k_n(R))^2}{2} = \frac{1}{R} -\frac{1}{2n^2}$.
 The charge distribution at the perturber is
 \be
 \label{eq:charge}
 \langle Q_L(R)\rangle_n = -\frac{(\frac{d \nu}{d E})^{-1}}{\pi k_n(R)}\frac{d\delta_L(\kappa)}{d \kappa}\bigg|_{\kappa=k_n(R)},
 \ee
 where $\nu$ is related to the electronic energy $E$ via  $E=-1/2\nu^2$. The presence of the Rydberg electron's density of states $\frac{d \nu}{d E}$ arises from the quantization of the $e-B$ scattering continuum by the Coulomb field of $A^*$ (for details see SM).
The charge in \cref{eq:charge} is proportional to the time-delay of the $e-B$ subsystem, which leads to a transparent interpretation \cite{smith1960lifetime}.
A large and positive time delay implies that the electron slows down near the perturber, which consequently appears as dressed with a negative charge.
A negative time delay has the opposite consequence: the electron spends less time near the perturber than elsewhere, dressing it with a positive charge.

The dressed anion and the positively charged Rydberg core interact via a Coulomb force $F_L(R) = \langle Q_L(R)\rangle_n/R^2$ yielding a potential energy $V_L(R)$ relative to the $n-$th manifold
\be
\label{eq:pec}
V_L(R) = \frac{1}{2n^2}-\frac{1}{2(n-\delta_L[k_n(R)]/\pi)^2},~~{\rm{for}}~R\le R_c,
\ee
where $V_L(R>R_c)=0$ for a perturber located outside the Rydberg's electron orbit ($R_c=2n^2$).
Note that \cref{eq:pec} was also obtained via different methods in Ref. \cite{BKmodel}, emphasizing its similar structure as the Rydberg formula. 
This highlights that the phase shifts play the role of molecular quantum defects \cite{DuGreene87}.
Due to the semiclassical nature of the dressed ion-pair model, \cref{eq:pec} captures only the shape of the molecular PECs, see orange lines in \cref{fig:fig1}(a).

\begin{figure}[t]
\includegraphics[width=0.98\columnwidth]{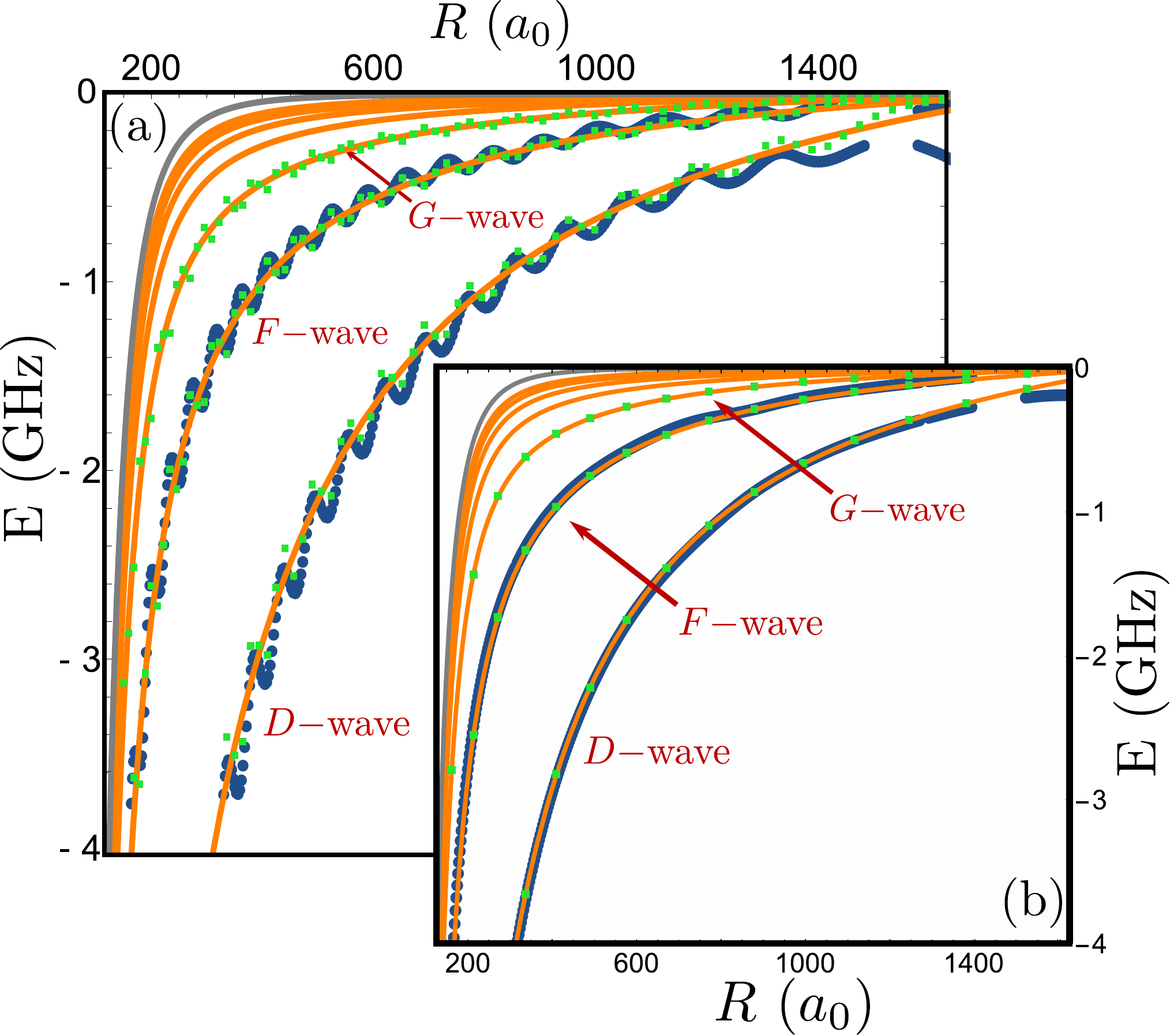}
\caption{(color online) The $L\ge 2$ PECs relative to the $n=30$ manifold for (a) $\Sigma$ and (b) $\Delta$ molecular symmetry using the BA phase shifts.
	The orange (blue) lines indicate the PECs within the ion-pair model (GLFT approach).
	The green squares denote the results of the numerical perturbation theory for a soft-core polarization potential.
	The polarization potential is shown in gray.
We have set $\mu_{l}=0$ for this comparison.
}
\label{fig:fig2}
\end{figure}

The effective charge in \cref{eq:charge} elucidates the emergence of Coulombic molecular PECs.
For a generic phase shift $\delta_L$, as for $S$ or $P$ partial waves, the charge fluctuates dramatically as $R$ changes, yielding non-Coulombic PECs.
However, at low-energies and for $L\ge2$ only the centrifugal barrier and the tail of the polarization potential contribute to the phase shifts, imparting on them a universal {\it linear} energy dependence.
Namely, within the Born approximation (BA) the $L\ge 2$ phase shift is
\begin{equation}
	\delta_L \approx \pi \overline{\alpha}_Lk^2	;~ ~\overline{\alpha}_L = \frac{\alpha}{(4L^2-1)(2L+3)},
\end{equation}
where we confirmed that this matches closely the calculated phase shifts for the alkali atoms (see SM).
The linear energy dependence yields an effective charge virtually independent of $R$ which imprints the Coulombic character onto the corresponding PECs.
Indeed, substituting the BA phase shifts in \cref{eq:pec} and expanding it in powers of $n^{-1}$ yields the corresponding Coulomb potential and additional higher order terms:
\be
\label{eq:rydberg}
U_L(R) = - \frac{\alpha}{2R^4}+\frac{\overline{\alpha}_L}{n^5}-\frac{2\overline{\alpha}_L}{n^3R}-\frac{6\overline{\alpha}_L^2}{n^4R^2}+\dots,
\ee
where in $U_L(R)$ the ion-atom polarization potential $-\frac{\alpha}{2R^4}$ is added.
The prefactor of the Coulomb term matches \cref{eq:charge} in the large $n$ limit where $\nu\to n$, i.e. $\braket{Q_L(R)}_n=2 \bar{\alpha}_L/n^3$, and for Rb the $L=2$ fractional charge is $\sim6.08/n^3$.
Considering only the dominant Coulomb and constant terms in \cref{eq:rydberg} results in a nuclear vibrational spectrum which obeys a Rydberg formula $E_{vJ}^{nL}$ with a redefined Rydberg constant $R_{nL}'$,
\be
\label{eq:qdef}
E_{vJ}^{nL} = \frac{\overline{\alpha}_L}{n^5} -\frac{R_{nL}'}{(v-\eta_J)^2};\,\,R_{nL}' = \frac{2m_{AB}\overline{\alpha}_L^2}{n^6}. 
\ee
$m_{AB}$ is the reduced mass of the molecule and $J$ is the nuclear angular momentum.
The constant energy shift,$\frac{\overline{\alpha}_L}{n^5}$, included in this formula is consistent with the truncation of the Rydberg series to a finite number of states by the vanishing of $U_L(R)$ at the classical turning point.
The maximum number of states is given by $v_\text{max} \approx \sqrt{2\overline{\alpha}_Lm_{AB}/n}$.
The nuclear quantum defect $\eta_J$ accounts for the effects of the non-Coulombic terms in \cref{eq:rydberg} as well as the complicated molecular potential energy curves at short internuclear distances, $R<30$~$a_0$.
 The nuclear defects $\eta_J$ are system-dependent and their evaluation is beyond the scope of this study.

 \cref{fig:fig2} shows the family of $L\ge2$ PECs relative to $n=30$ manifold for (a) $\Sigma$ ($M=0$) and (b) $\Delta$ ($M=2$) molecular symmetry, where for simplicity we use the BA phase shifts and set $\mu_l =0$.
In both panels the PECs within the dressed ion-pair model (orange) condense to the atom-ion potential (gray) as $L\to \infty$.
As seen in butterfly ULRMs, the PEC oscillations disappear with increasing $M$ \cite{HamiltonGreeneSadeghpour,KhuskivadzeJPB,EilesButterflyTrimer}.
 Panel (b) shows the smooth $\Delta$ PECs: the results of the GLFT method (blue) are in excellent agreement with the dressed ion-pair model (orange).

 In general, the calculations of GLFT theory for $L\le1$ match those obtained by the Green's function method \cite{Crowell,HamiltonThesis,KhuskivadzePRA} and Omont's pseudopotential (see SM for details).
Note that, these methods share a crucial approximation: the phase shifts are fully accumulated at the perturber, and thus the polarization potential is replaced with a zero-range $e-B$ interaction.
 The validity of this approximation breaks down near the classical turning point where the $e-B$ momentum vanishes yielding divergent scattering volumes $-\tan \delta_L(k)/k^{2L+1}$ for $L\ge1$~\cite{du_PRA_1987a}.
 This low-energy unphysical behavior can in principle invalidate the molecular PECs especially for $L\ge 2$.
 Therefore, the $L\ge 2$ PECs were numerically calculated by diagonalizing a divergence-free soft-core polarization potential $V_{e-B}(\boldsymbol{r}) =-\frac{\alpha}{2}(\beta^4+|\boldsymbol{r}-\boldsymbol{R}|^4)^{-1}$ avoiding the use of the phase shifts altogether.
In \cref{fig:fig2} the soft-core results (green squares) match the GLFT results (orange curves) for all $R<R_c$, beyond which the GLFT, and in fact all methods using phase shifts as external input, break down.

\begin{figure}[t]
	\includegraphics[width=0.9\columnwidth]{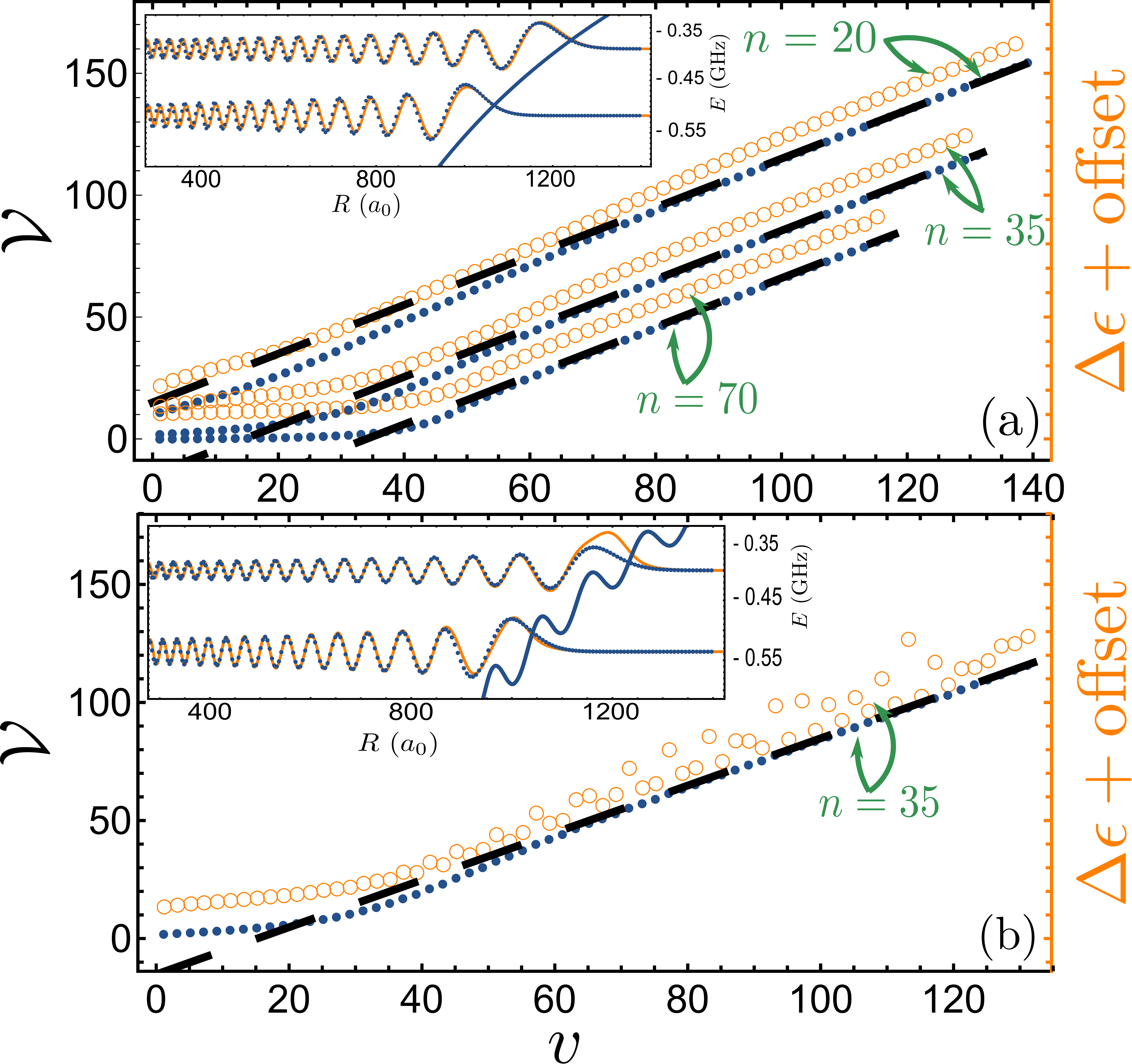}
	\caption{(color online) The effective nuclear quantum number $\mathcal{V}$ (blue dots) and the rescaled difference of successive energy levels $\Delta \epsilon$ (orange circles) for the (a) $\Delta$ and (b) $\Sigma$ ''dragonfly'' molecular curves at different $n-$manifolds.
	The dashed lines correspond to linear fits to $\mathcal{V}$ with unity slope.
The insets depict the nuclear wavefunction at two different eigenenergies evaluated numerically (orange line) or the fitted Coulomb wavefunctions (blue dots). Note that an arbitrary offset is added to $\Delta \epsilon$ for illustration purposes.}
\label{fig:fig3}
\end{figure}

\cref{fig:fig3} confirms the Rydberg character of the vibrational spectrum of the adiabatic Born-Oppenheimer dragonfly PECs, which we focus on as they are the deepest of the $L\ge 2$ PECs.
Here, we remove any approximations on the phase shifts and use the numerical, rather than the BA, $\delta_L$. 
We show results for fixed nuclear rotational quantum number $J= 0$.
The spectra, $\varepsilon_{vJ}^{nL}$, are obtained numerically using a hard wall at $R=30~a_0$ that mimics the short-range physics.
Here, a  more detailed theoretical description of the latter would only result in modified quantum defects $\eta_J$, maintaining the Rydberg characteristics of the $\varepsilon_{vJ}^{nL}$.

Across several $n$-manifolds, \cref{fig:fig3}(a) shows the effective nuclear quantum number $\mathcal{V}=\sqrt{R'_{nL}/(\bar{\alpha}_L/n^5-\varepsilon_{vJ}^{nL})}$ (blue dots) for the $\Delta$ dragonfly PECs.
The black dashed lines show the $\mathcal{V} = v-\eta_J$, derived from \cref{eq:qdef}, where we have fitted $\eta_J$ to the numerical data.
The evident linear dependence manifests a tHRS.
For high $n-$ manifolds, i.e. $n=70$, the nuclear states for $v>50$ yield a Rydberg spectrum; deviations occur at $v<50$ due to the polarization potential.
At low $n-$manifolds, i.e. $n=20$, nearly all the nuclear states form a Rydberg series since the Coulomb-like PEC dominates the polarization potential due to the relatively large effective charge.
In general, at large $v$, the quantum numbers $\mathcal{V}$ agree well with the dashed black lines which correspond to \cref{eq:qdef} with a fitted nuclear quantum defect $\eta_J$.

\cref{fig:fig3}(a) also shows the rescaled difference of successive energy levels, $\Delta \epsilon = \left|2(\varepsilon_{vJ}^{nL}-\varepsilon_{v-1J}^{nL})/R_{nL}'\right|^{-1/3}$ (orange circles), which more sensitively probes the Rydberg character of the series.
The linear dependence of $\Delta \epsilon$ on $v$ with unity slope indicates that the nuclear states possess the proper Rydberg energy scaling, i.e $\Delta E\sim1/v^3$.

\cref{fig:fig3}(b) shows the same quantities as in panel (a) for the $\Sigma$ ``dragonfly'' PEC.
In \cref{fig:fig2}(a) the potential wells are too shallow to support localized bound vibrational states.
Instead, they produce low-amplitude oscillations of $\mathcal{V}$ around the linear fit (black dashed lines) which are seen faintly on this scale.
The $\Delta\epsilon$ values show this modulation more explicitly, highlighting the non-Rydberg nature of the corresponding spectrum.
The insets in \cref{fig:fig3} illustrate the numerical nuclear wavefunctions (orange lines) at two different eigenenergies; The blue dots denote fits to the Coulomb wave functions.
Evidently, in panel (a) the numerical and the fitted wavefunctions are in excellent agreement.
In panel (b) deviations in the outermost lobes are observed due to the wells in the $\Sigma $ PEC, but the overall nodal pattern of the wave functions is determined by the dominant energy scale of the Coulomb-like potential.

The laboratory excitation of tHRS will be similar to that of trilobite-like ULRMs, requiring three-photon excitation via admixture of the $nf$ quantum defect state of Rb or by using other, more favorable, quantum defect states in other atoms \cite{EilesHetero,TallantCS,Eiles2015}.
The nuclear quantum defects can be extracted by scanning the appropriate energy range to obtain a spectrum which can be fit to the Rydberg formula.
The short-range physics at $R<30~a_0$, where molecular ion formation and other ultracold chemistry can occur, is beyond the scope of the present work, but it will affect the size of the nuclear quantum defects and, more importantly, the lifetime of the Rydberg states \cite{UltracoldChem}.
A firm upper limit on the lifetime is set by the electronic Rydberg lifetime, which depends on $(n,~l)$ quantum numbers and for the $n$ values considered in typical experiments can range from $1-100\mu$s.
However, as in the HRS, the lifetimes will likely be substantially reduced by nuclear decay channels to $\sim1-100$ns.
These lifetimes increase with both $v$ and $J$; $J$ could be increased by applying a very weak electric field to create pendular states \cite{Rost}.

In summary, we have identified new vibrational states in ultralong-range molecules which form a trimmed heavy Rydberg series with very small Rydberg constant.
A generalized GLFT method enables us to accurately determine the underlying highly excited Coulomb-like PECs that stem from the interaction with the perturber.
Although we focused on these new attributes, the effects of higher partial waves may contribute to more accurate theoretical spectra to compare with ongoing experimental efforts \cite{MacLennan,PhasesFeyMeinert,SpinFey,EilesSpin,EilesHetero}.
Finally, because of the generic character of the presented binding mechanism, we expect that similar trimmed Rydberg series, as described and analyzed here in the context of ULRMs, could occur in any system containing a Rydberg atom and a polarizable perturber, e.g. atoms with more complex structure, multiple atoms, excitons, or even large compounds like fullerines or nano-droplets.

\begin{acknowledgments}
We thank Chris H. Greene for useful comments and suggestions about the manuscript.
MTE is grateful to the Alexander von Humboldt stiftung for support through a postdoctoral fellowship.
FR acknowledges the financial support by the U.S. Department of Energy (DOE), Office of Science, Basic Energy Sciences (BES) under Award No. DE-SC0012193.
The numerical calculations have been performed using NSF XSEDE Resource Allocation No. TG-PHY150003.
\end{acknowledgments}
\bibliography{Thesis_bib.bib}

\clearpage
\onecolumngrid
\begin{center}
\textbf{\large Supplemental Material: Dressed ion-pair states of an ultralong-range Rydberg molecule}
\end{center}
\twocolumngrid
\setcounter{equation}{0}
\setcounter{figure}{0}
\setcounter{table}{0}
\setcounter{page}{1}
\makeatletter
\renewcommand{\theequation}{S\arabic{equation}}
\renewcommand{\thefigure}{S\arabic{figure}}

In this supplementary material additional information is provided on various aspects of the main manuscript.
In summary, these extra details cover the following points:
\begin{itemize}
	\item Alkali atom-electron phase shifts for high angular momentum $L$.
	\item The dipole moments of the dragonfly molecules and the $L\ge2$ molecular wavefunctions for all the molecular symmetries $M$.
	\item Derivations related to the Omont pseudopotential for the dragonfly potential curve and its comparison to the generalized local frame transformation theory (GLFT).
	\item Details on the derivations of the dressed ion-pair model.
\end{itemize}

\section{Phase shifts for $L\ge 2$}
\begin{figure}[b!]
\includegraphics[width=\columnwidth]{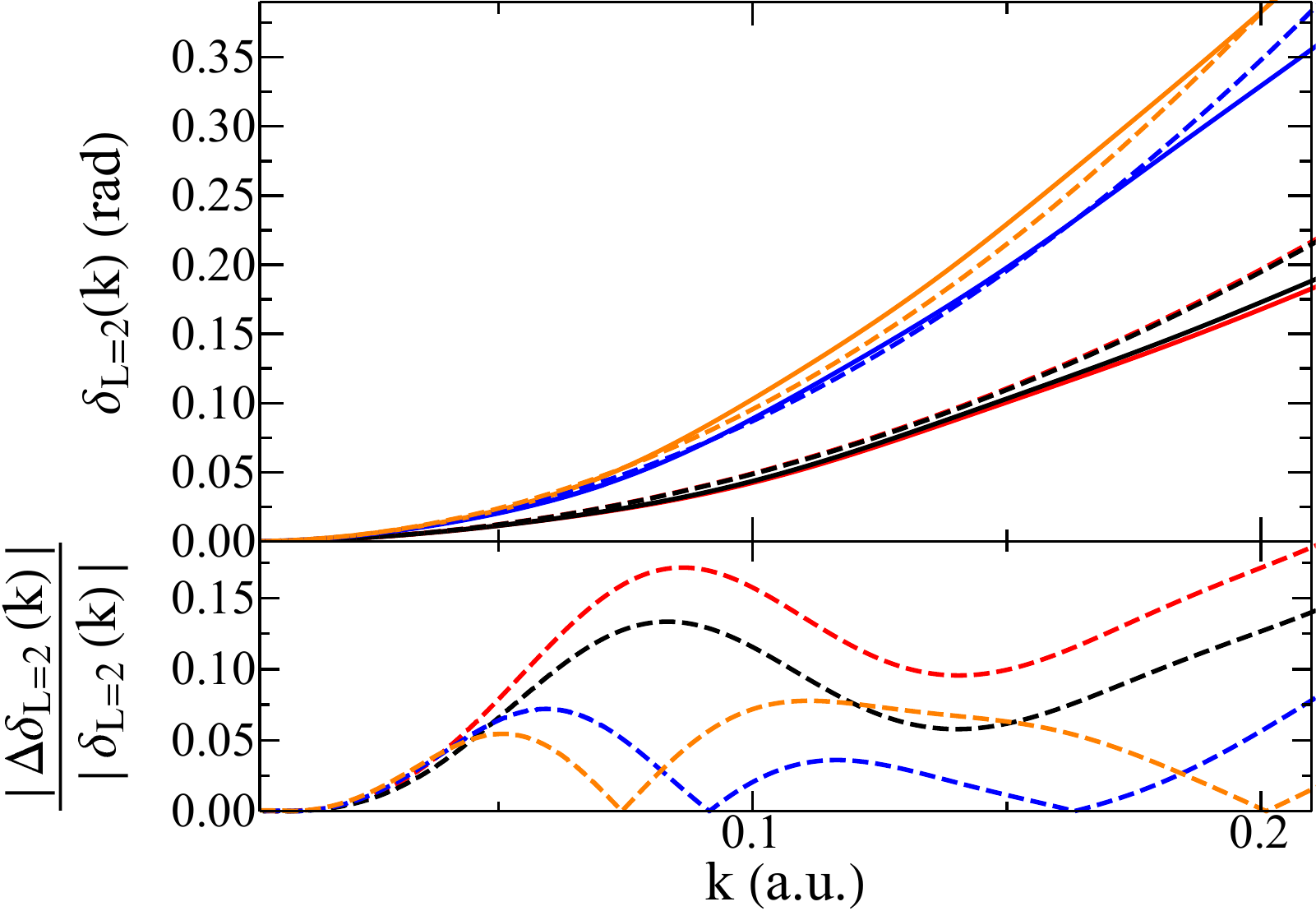}
\caption{\label{fig:phases}(color online) The four triplet $D$ phase shifts for Li, Na, K, and Rb (red, black, blue, and orange, respectively) computed using the $R$-matrix code (solid) and the Born approximation (dashed lines). The polarizabilities are 164.2 $a_0$, 162.7 $a_0$, 290.6 $a_0$, and 319.2 $a_0$ for Li, Na, K, and Rb respectively. The bottom panel shows the relative error between the exact and Born approximate phase shifts.}
\end{figure}
Using a model potential for the neutral atom \cite{Marinescu} and the $R$-matrix method to compute the logarithmic derivative of the scattering wave function at the surface of the volume where electron-electron correlation and exchange are relevant, we computed the non-relativistic phase shifts for $D$ and $F$ waves. Outside of the two-electron region the appropriate asymptotic solutions were obtained by numerically propagating the wave function using a Numerov algorithm in the polarization potential. This method was used in Ref. \cite{EilesHetero} to compute $S$ and $P$-wave phase shifts and was shown to be in excellent agreement with existing highly accurate theoretical calculations, and additionally we confirmed that our lithium $D$-wave phase shift results agreed with the recent calculation of Tarana and Curik \cite{TaranaCurikLi}.
In \cref{fig:phases}(a) we show the results of this calculation for the $D$-wave phase shifts of Li, Na, K, and Rb (the relativistic effects in Cs are too large to make a non-relativistic calculation reliable), along with the Born approximation for the phase shifts.
The relative error between the exact calculation and the Born approximated phase shifts in \cref{fig:phases}(b) reveals that all phase shifts are within 15\% of the $k^2$ form of the Born-approximation all the way to high momenta, i.e. $k\sim 0.2~(a.u.)$. We note that potassium fits best to the Born approximation at the highest energies relevant to ultralong-range Rydberg molecules (ULRMs).

\section{Dipole moments and Ultra-long range molecular wavefunctions with different symmetries}

\begin{figure}[t!]
\includegraphics[width=\columnwidth]{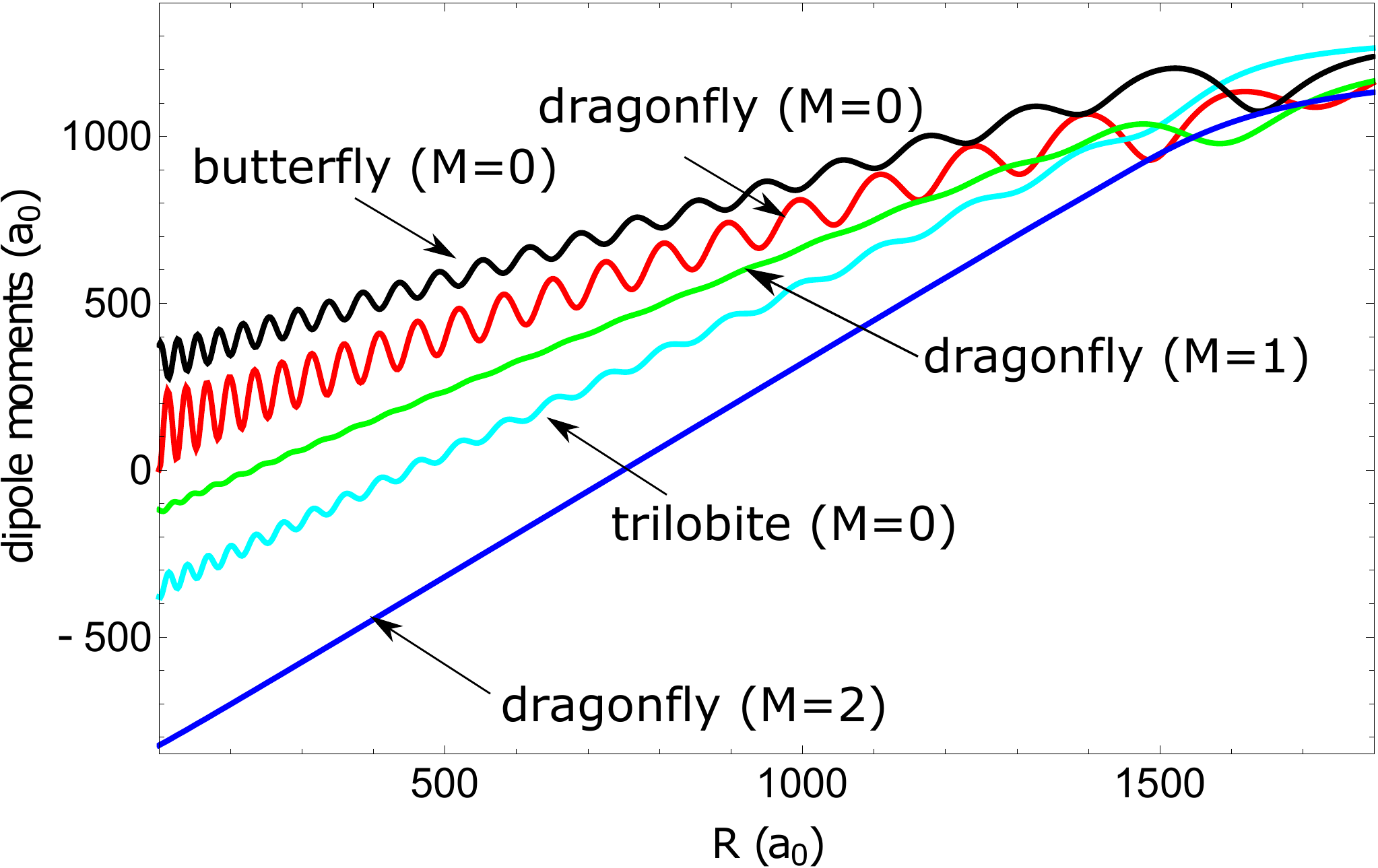}
\caption{\label{fig:s1}(color online) The dipole moments for the dragonfly ultralong-range  Rydberg molecules shown for different molecular symmetries $M$ using Rb atoms at the $n=30$ electronic Rydberg manifold. The dipole moments of the trilobite and butterfly molecules are also shown for comparison with the dragonfly ones.}
\end{figure}

\begin{figure*}[ht!]
\includegraphics[width=\textwidth]{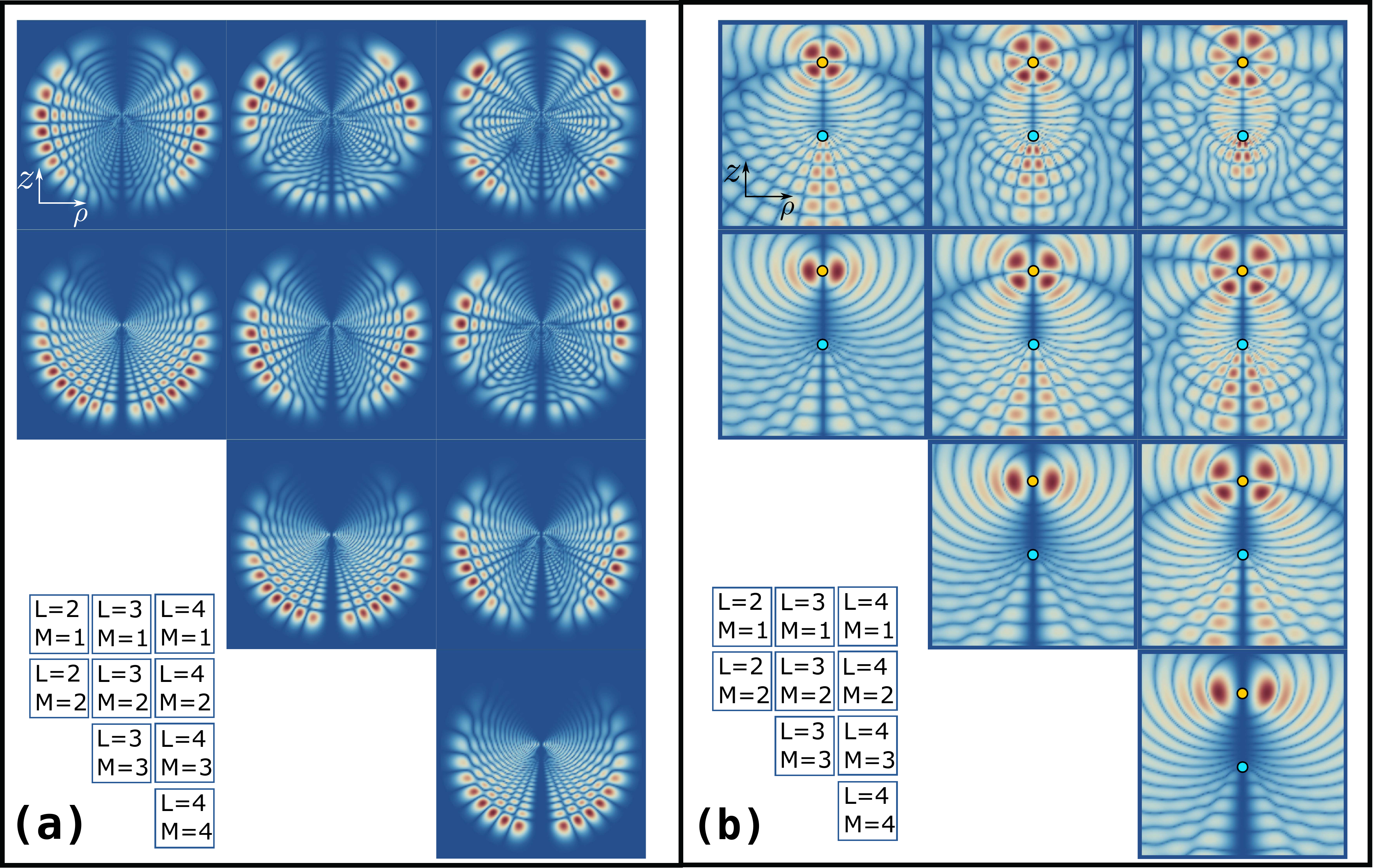}
\caption{\label{fig:s2}(color online) The wavefunctions for the dragonfly, firefly and gadfly Rb$_2$ molecules tabulated with respect to the orbital angular momentum $L$ (columns) and the corresponding projection quantum number $M$ (rows). Panel (a) depicts the wavefunction within entire Rydberg's electron orbit whereas panel (b) show zoom in plots close to the perturber. The blue (orange) dot indicates the location of the Rydberg core (perturber).
	Note that panel (a) shows $\rho\sqrt{| \Psi_M(z,\rho)|}$ for $z \in[-2000,2000]$ and $\rho \in[-2000,2000]$.
	Panel (b) shows $\sqrt{|\Psi_M(z,\rho)|}$ for $z \in[-250,300]$ and $\rho \in[-250,250]$.
}
\end{figure*}
As mentioned in the main text the high $L$ ULRMs have dipole moments on the same strength as in the case of the butterfly molecules.
\cref{fig:s1} depicts the dipole moments for the dragonfly molecules with different molecular symmetry, namely $\Sigma$ (red), $\Pi$ (green) and $\Delta$ (blue). 
 \cref{fig:s1} also shows the dipole moments of the trilobite (black line) and butterfly (light blue line) molecules for $\Sigma$ molecular symmetry.
It is evident by this comparison in \cref{fig:s1} that the dragonfly molecules exhibit dipole moments that have the same qualitative features as the butterfly ones.

In  Fig.1(b) in the main text we presented the $\Sigma$ ($M=0$) ULRM molecular wavefunctions for the dragonfly ($L=2$), firefly ($L=3$) and gadfly ($L=4$) potential curves.
In \cref{fig:s2} we provide the complete table of the wavefunctions for all the molecular symmetries.  Panel (a) depicts the wavefunctions over the entire Rydberg's electron orbit.
Panel (b) consists of a zoom in of the wavefunctions in panel (a) near the location of the perturber.
As mentioned in the main text we observe in \cref{fig:s2}(b) that the molecular wavefunction exhibits around the perturber the symmetry as the electron-perturber subsystem.
In addition, we observe that along the diagonals of the tables in \cref{fig:s2} the molecular wavefunctions qualitatively are the same, manifesting locally at the perturber the same nodal structure.
The nodal pattern for $L\ge1$ and $M>0$ obeys the rule ${\rm{nodes}}=1+L-M$.

\section{Omont's peudopotential method for dragonfly potential curves}
To confirm the accuracy of the GLFT method, and to provide an alternative method closer to the typical approach in the field of Rydberg molecules, we have derived also the expressions for the matrix elements of the $D$-wave component of the Omont pseudopotential. This pseudopotential, although quite simple for $s$- and $p-$wave scattering, becomes increasingly tedious to evaluate as $L$ increases. The $D$-wave operator reads
\begin{align}
	V_\text{D}(\vec r) &= 10\pi \delta^3(\vec r - \vec R)\left(-\frac{\tan\delta_2[k(R)]}{k(R)}\right)P_2\left(\frac{\cev\nabla\cdot\vec\nabla}{[k(R)]^2}\right),
\end{align}
where $P_2$ is the Legendre polynomial of second order, and $\vec\nabla$ is a gradient operator acting on the wave function in the direction of the arrow when determining matrix elements. We next construct the matrix element $V_{\alpha\alpha'}^\text{D}$, where $\alpha$ is a collective quantum number standing in for $n$ (principal quantum number), $l$ (orbital angular momentum), and $m$ (azimuthal quantum number). We use the Schr\"odinger equation, $\nabla^2\Psi_{nlm}(\vec R) = -k^2\Psi_{nlm}(\vec R)$, to factor out $k(R)$ terms such that the $D$-wave scattering ``volume'', $a_d^5[k(R)] = -\frac{\tan\left(\delta_2[k(R)]\right)}{[k(R)]^5}$, can be defined.
Under these considerations, the matrix elements $V_{\alpha\alpha'}^\text{D}$ read:
\begin{align}
	V_{\alpha\alpha'}^\text{D} &= 5\pi a_d^5[k(R)]\bra{nlm}\left(3\left(\cev\nabla\cdot\vec\nabla\right)^2-\cev\nabla^2\vec\nabla^2\right)\ket{n'l'm'}.
\end{align}

The simplest pathway to evaluate these matrix elements is to evaluate the gradient operators in cartesian coordinates to avoid difficulties with the non-commutivity of the spherical operators, and then to transform back into spherical coordinates in the final step. Due to the cylindrical symmetry, this can be done for each value of $m = m' = M$ individually.
\begin{itemize}
		
	\item For  $M = 0$ we have the following expression:
\begin{widetext}
\begin{align}
V_{\alpha\alpha'}^{\text{D},M=0} &= -\frac{5 a_d^5}{8R^6}\left(f_{nl}(R)u_{nl}(R)-6R u_{nl}'(R)\right)\left(f_{nl'}(R)u_{nl'}(R)-6R u_{nl'}'(R)\right)\sqrt{(2l+1)(2l'+1)}
\end{align}
\end{widetext}
where
\be
 f_{nl}(R) = 6 + 3l(l+1) - 4R+2(R/n)^2
\ee
and $u_{nl}'(r) = \frac{d u_{nl}(r)}{dr}$. 
Note that this expression assumes $\vec R = R\hat z$ and $m = 0$.

\item For the $M = 1$ case we obtain the following relation:
\begin{widetext}
\begin{align}
 \bra{nl1}V_\text{dwave}\ket{n'l'1}&= -\frac{15 a_d^5[k(R)]}{4R^6}\sqrt{(2l+1)(l+1)l(2l'+1)(l'+1)l'}\left[Ru'_{nl}(R)-2u_{nl}(R)\right]\left[Ru'_{nl'}(R)-2u_{nl'}(R)\right]
\end{align}
\end{widetext}
\item And finally, the matrix elements for the $M = 2$ case read:

\begin{align}
 \bra{nl2}V_\text{dwave}\ket{n'l'2}&= -\frac{15 a_d^5[k(R)]}{16R^6}g_{l}g_{l'}u_{nl}(R)u_{nl'}(R),\end{align}
 where the terms $g_l$ obey the relation $g_{l}=\sqrt{(2l+1)(l+2)(l+1)l(l-1)}$.
\end{itemize}

\begin{figure}[t!]
\includegraphics[width=\columnwidth]{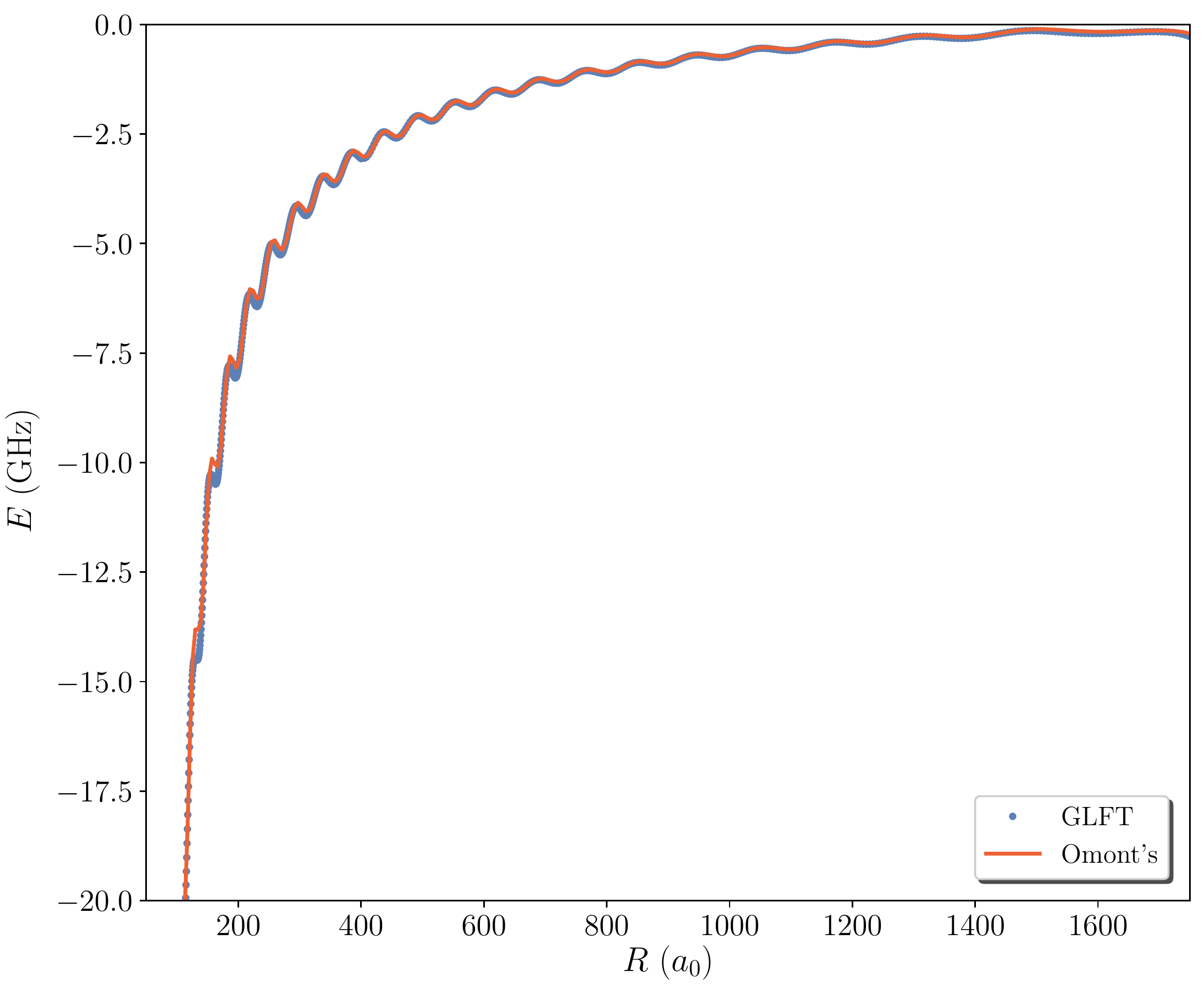}
\caption{\label{fig:somont}(color online) Comparison between the GLFT (blue dots) and Omont's pseudopotential (red line) approach for the dragonfly molecular curve for Rb atoms at $n=30$ electronic Rydberg manifold. Note that the atomic quantum defects $\mu_l$ are set equal to zero.}
\end{figure}

\cref{fig:somont} illustrates a comparison between the standard Omont's pseudopotential theory (red line) and the GLFT approach (blue dots) only for the $\Sigma$ dragonfly potential energy curve for Rb atoms at the $n=30$ electronic Rydberg manifold.
Note that for simplicity the atomic quantum defects $\mu_l$ are neglected.
Evidently, \cref{fig:somont} highlights that the GLFT treatment is in excellent agreement with the standard techniques used in the field of Rydberg molecules.

\section{Dressed ion-pair model: details and derivations}
In the following, additional details on the dressed ion-pair model are provided.
The key quantity is the excess probability of the Rydberg electron in the near vicinity of the perturber which is defined as the difference of the electron's probability with and without the presence of the perturber.
The spatial volume integral of electron's excess probability is interpreted as a charge distribution that dresses the neutral atom, i.e. the perturber.
Assuming a sphere of radius $\xi_0$ around the perturber, the charge distribution around the perturber is written in terms of the excess probability according to the following expression:
  \be
  Q_L(\xi_0,R) =-\int_{\xi\le \xi_0} \left(|\Phi_L(\boldsymbol{\xi})|^2 - |\Phi_L^0(\boldsymbol{\xi})|^2\right)\dd{\boldsymbol{\xi}},
  \label{eq:xi0charge}
  \ee
where  $\Phi_L(\xi)$ and $\Phi_L^0(\xi)$ are the electron-perturber scattered / un-scattered wavefunctions, respectively.
For large enough radius $\xi_0$ such that the electron's wavefunctions behaves as
\begin{equation}
	\Phi_L(\boldsymbol{\xi}_0)\to A\sqrt{\frac{2}{\pi k}} \sin[k \xi_0-L \pi/2-\delta_L(k)]Y_{LM}(\hat{\xi}_0),
	\label{eq:scatwfn}
\end{equation}
where $\delta_L(k)$ is the electron-perturber scattering phase shift and $A=\nu^{-3/2}$.
Note that by setting $\delta_L(k)=0$ \cref{eq:scatwfn} provides us with the $\Phi_L^0(\boldsymbol{\xi})$ wavefunction.

The constant $A$ is defined such that $\int \dd \boldsymbol{\xi }|\Phi_L(\boldsymbol{\xi})|^2=\delta_{\nu \nu'}$ where $\nu=1/\sqrt{-2 E}$ is the effective principal quantum number.
In essence, the constant $A$ takes into account that the electron-perturber scattering takes place in the presence of the Coulomb field generated by the residual core and the Rydberg electron.
Thus, the electron-perturber collisional energy is discretized by the Coulomb interaction.
This means that $A^2$ gives the inverse of the Rydberg electron's density of states, i.e.  $A^2\equiv (\frac{d \nu}{d E})^{-1}$.

Plugging \cref{eq:scatwfn} into \cref{eq:xi0charge} and converting the volume integral into a surface one we obtain a closed form expression for the charge distribution.
\begin{align}
Q_L(\xi_0, R)= &-\bigg(\frac{d \nu}{d E}\bigg)^{-1} \bigg(\frac{\delta'_L(k)}{ \pi k} \nonumber \\
&-\frac{\cos(2k \xi_0-L \pi-\delta_L(k))\sin \delta_L(k)}{ \pi k^2}\bigg),
	\label{eq:surfxi0charge}
\end{align}
where $\delta'_L(k)=\frac{d \delta_L(k)}{d k}$.
In the preceding equation the charge distribution depends on the radius of the sphere of integration $\xi_0$ exhibiting an oscillatory behavior as $\xi_0$ increases.
However, we are interested in the charge distribution only around the perturber.
In the spirit of Smith in Ref.\cite{smith1960lifetime}, this oscillatory behavior can be eliminated  by averaging over a cycle $\xi_0$ and take the limit $\xi_0\to\infty$  yielding the expression:
\begin{equation}
	\braket{Q_L(R)}=\lim_{\xi_0\to\infty}\frac{1}{\xi_0}\int_{0}^{ \xi_0}\dd x Q_L(x,R)=-\bigg(\frac{d \nu}{d E}\bigg)^{-1} \frac{\delta'_L(k)}{\pi k}
	\label{eq:scharge}
\end{equation}

In order to obtain the potential curves via \cref{eq:scharge} in a simple form we approximate the momentum as $k(R)\equiv k_n(R)=\sqrt{2/R-1/n^2}$ as is usually done in the diagonalization approach for Rydberg molecules.
This implies that we define the electron-perturber momentum relative to the $ n$ hydrogenic Rydberg manifold.
Using this approximation in \cref{eq:scharge} we obtain the relation used in the main text
\begin{equation}
	\braket{Q_L(R)}_n=-\frac{(\frac{d \nu}{d E})^{-1} }{ \pi k_n(R)}\frac{d \delta_L(\kappa)}{d\kappa}\bigg|_{\kappa=k_n(R)}.
\end{equation}
Note that the quantity $\frac{1}{\pi \kappa}\frac{d \delta_L(\kappa)}{d\kappa}$ for $\kappa=k_n(R)$ is the time-delay for half-collisions \cite{smith1960lifetime}.
\end{document}